\documentclass[english,aps,prl,twocolumn,amsmath,amssymb,showpacs,superscriptaddress,notitlepage,longbibliography]{revtex4-2}

\usepackage[T1]{fontenc}
\usepackage{color}
\definecolor{page_backgroundcolor}{rgb}{1, 1, 1}
\pagecolor{page_backgroundcolor}
\usepackage{amsmath}
\usepackage{graphicx}
\usepackage{microtype}
\usepackage{bm}
\usepackage{makecell}
\usepackage[normalem]{ulem}
\usepackage{enumitem}
\usepackage[marginal]{footmisc}
\usepackage{diagbox}

\makeatletter
\usepackage{amssymb}
\usepackage{graphicx}
\usepackage[colorlinks=true,linkcolor=blue,anchorcolor=red,citecolor=blue,urlcolor=blue]{hyperref}
\usepackage[caption=false]{subfig}
\usepackage{lipsum}

\makeatother

\usepackage{babel}
\renewcommand{\appendixname}{\MakeUppercase{\appendixname}}
\begin{document}
\global\long\def\figurename{Fig.}

\title{Nonreciprocal Superconducting Transport from Chiral Edge States}

\author{Jin-Xing Hou }
\thanks{These authors contributed equally.}
\affiliation{Hefei National Laboratory, Hefei, 230088, China}

\author{Yan-Song Song}
\thanks{These authors contributed equally.}

\affiliation{School of Emerging Technology, University of Science and Technology of China, Hefei, 230026, China}
\affiliation{Hefei National Laboratory, Hefei, 230088, China}

\author{James Jun He}
\affiliation{Hefei National Laboratory, Hefei, 230088, China}
\affiliation{School of Emerging Technology, University of Science and Technology of China, Hefei, 230026, China}

 \author{Song-Bo Zhang}
 \email{Corresponding author: songbozhang@ustc.edu.cn}
 \affiliation{Hefei National Laboratory, Hefei, 230088, China}
 \affiliation{School of Emerging Technology, University of Science and Technology of China, Hefei, 230026, China}

\date{\today}

\begin{abstract}

Nonreciprocal superconducting transport enables dissipationless rectification and has attracted considerable interest, yet its microscopic origin is typically sought in bulk electronic states. Here, we show that boundary-controlled chiral edge states in topological systems provide a simple yet largely overlooked mechanism for nonreciprocal superconducting transport. Focusing on chiral kagome antiferromagnets, we demonstrate that out-of-plane spin canting or spin-orbit coupling opens a high-Chern-number bulk gap, giving rise to multiple chiral edge modes. Strikingly, sublattice-dependent boundary termination selects a single-valley character for the edge states, leading to asymmetric edge spectra at opposite edges. This boundary asymmetry directly yields observable nonreciprocal signatures in Josephson junctions oriented transverse to the edges, including asymmetric Andreev spectra, Josephson diode effect, and anomalous Fraunhofer interference patterns. 
These findings broaden the microscopic understanding of superconducting nonreciprocity and highlight boundary engineering as a tunable route toward superconducting diode devices.

\end{abstract}

\maketitle


\newpage
\noindent

\textit{\color{blue}Introduction}.\textemdash 
Directional control of charge transport is a major theme in quantum materials, underpinning rectification and nonreciprocal device functionalities~\cite{Rikken01PRL,Rikken05PRL,Ideue17Nphys,Tokura18NC}. In superconducting systems, this directionality manifests as the superconducting diode~\cite{Qin17NC,Hoshino18PRB,ando2020observation,lyu2021superconducting,HeJ2022NJP,Yuan2022PNAS,DaidoPRL2022,lin2022zero,Ilic22PRL,bauriedl2022supercurrent,HouPRL2023,nadeem2023superconducting,he2023supercurrent,le2024superconducting,BanerjeePRB2024,ChakrabortyPRL2025,wang2025universal,fracassi2025intrinsic,bhowmik2025optimizing,pal2025topological,bhowmik2025topological,li2026field,Wakatsuki17SA} and Josephson diode effects~\cite{wu2022Nature,Baumgartner22Nnanotech,Pal22Nphys,Hu2007PRL,Chen2018PRB,misaki2021theory,davydova2022universal,zhang2022general,tanaka2022theory,diez2023symmetry,hu2023josephson,Steiner2023PRL,sundaresh2023diamagnetic,Cheng2024PRB,PRB2024Superconducting,jiang2022superconducting,wang2025current,ma2025field,Sun2025PRB,sun2026strong,hou2025field,Lahiri2026PRB}, characterized by direction-dependent critical currents. Such effects typically require the breaking of both inversion and time-reversal symmetries. 
Existing mechanisms for this nonreciprocity are generally formulated in terms of asymmetries of bulk electronic structure, often arising from spin–orbit coupling together with Zeeman or exchange fields~\cite{DaidoPRL2022,HeJ2022NJP,Yuan2022PNAS}.
This naturally raises the question of whether superconducting nonreciprocity can originate from a different physical principle.

Topological chiral edge states provide a natural setting for addressing this question. Chiral edge transport and superconductivity represent two distinct paradigms of dissipationless transport. Although both have been extensively studied, their interplay in governing superconducting transport remains largely unexplored. 
In this work, we show that chiral edge states provide a simple yet largely overlooked mechanism for superconducting nonreciprocity through boundary engineering.

\begin{figure}[t]
\centering
\includegraphics[width=1\linewidth]{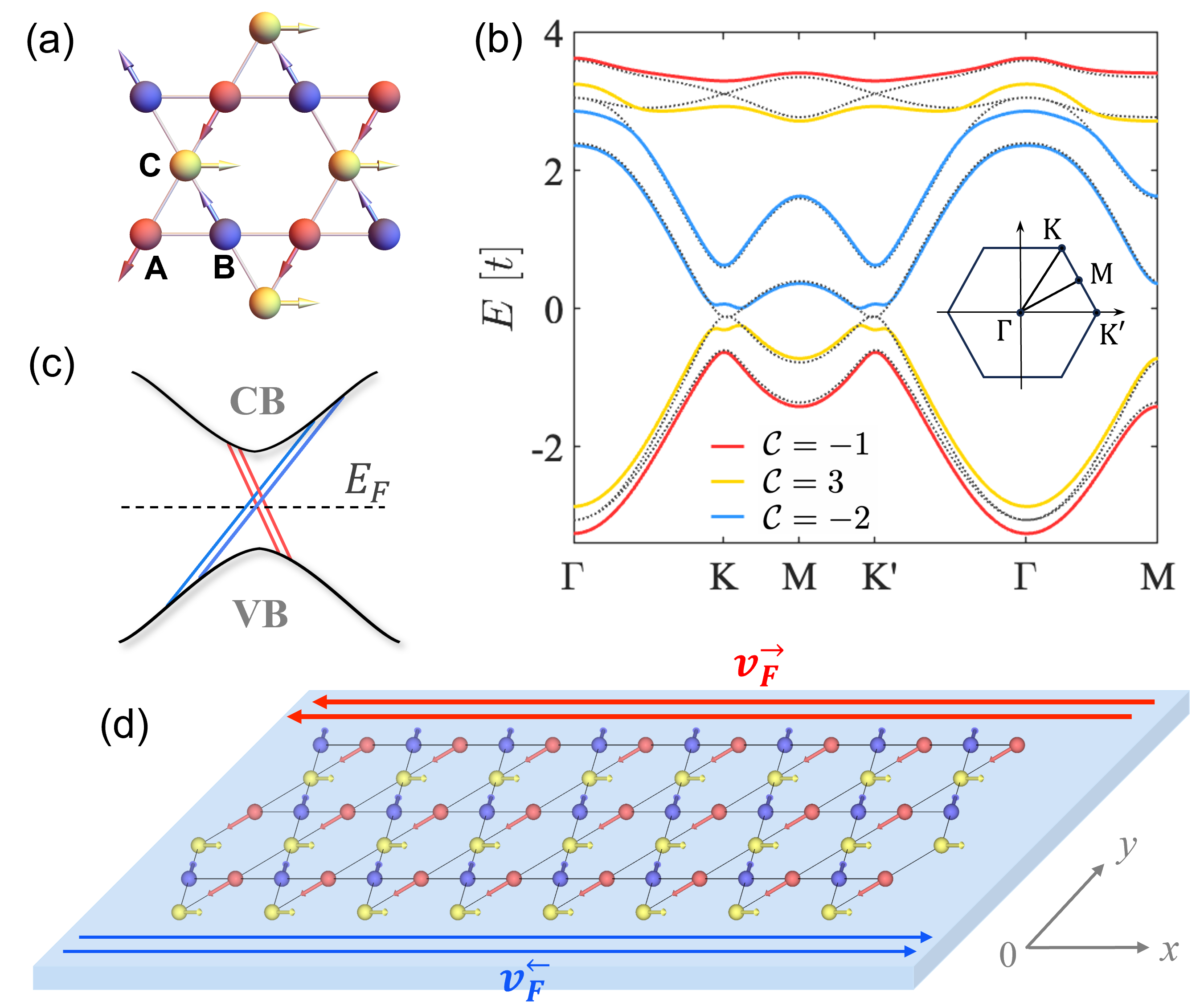}
\caption{(a) Schematic of the chiral kagome antiferromagnet with three sublattices ($A$, $B$, $C$) and coplanar $120^\circ$ magnetic order (colored arrows).  (b) Bulk band structure along high-symmetry lines of the Brillouin zone (inset) for $J=0.6t$ and $M_z=0.2t$. Bands are colored by their Chern numbers; dashed gray curves show the spectrum for $M_z=0$. An out-of-plane spin canting $M_z$ opens a bulk gap with nonzero Chern number. The reference energy is set at the Dirac point without magnetic order. (c) Schematic of chiral edge states traversing the bulk gap with inequivalent dispersions at opposite edges. Black curves denote the conduction (CB) and valence band (VB). (d) Ribbon geometry with an (A/B)-terminated upper edge and a (C)-terminated lower edge. The chiral edge channels propagate along opposite edges with different velocities $v_F^{\rightarrow/\leftarrow}$. 
}
\label{fig:Lattice_Band}
\end{figure}

\begin{figure*}[t]
\centering
\includegraphics[width=0.88\linewidth]{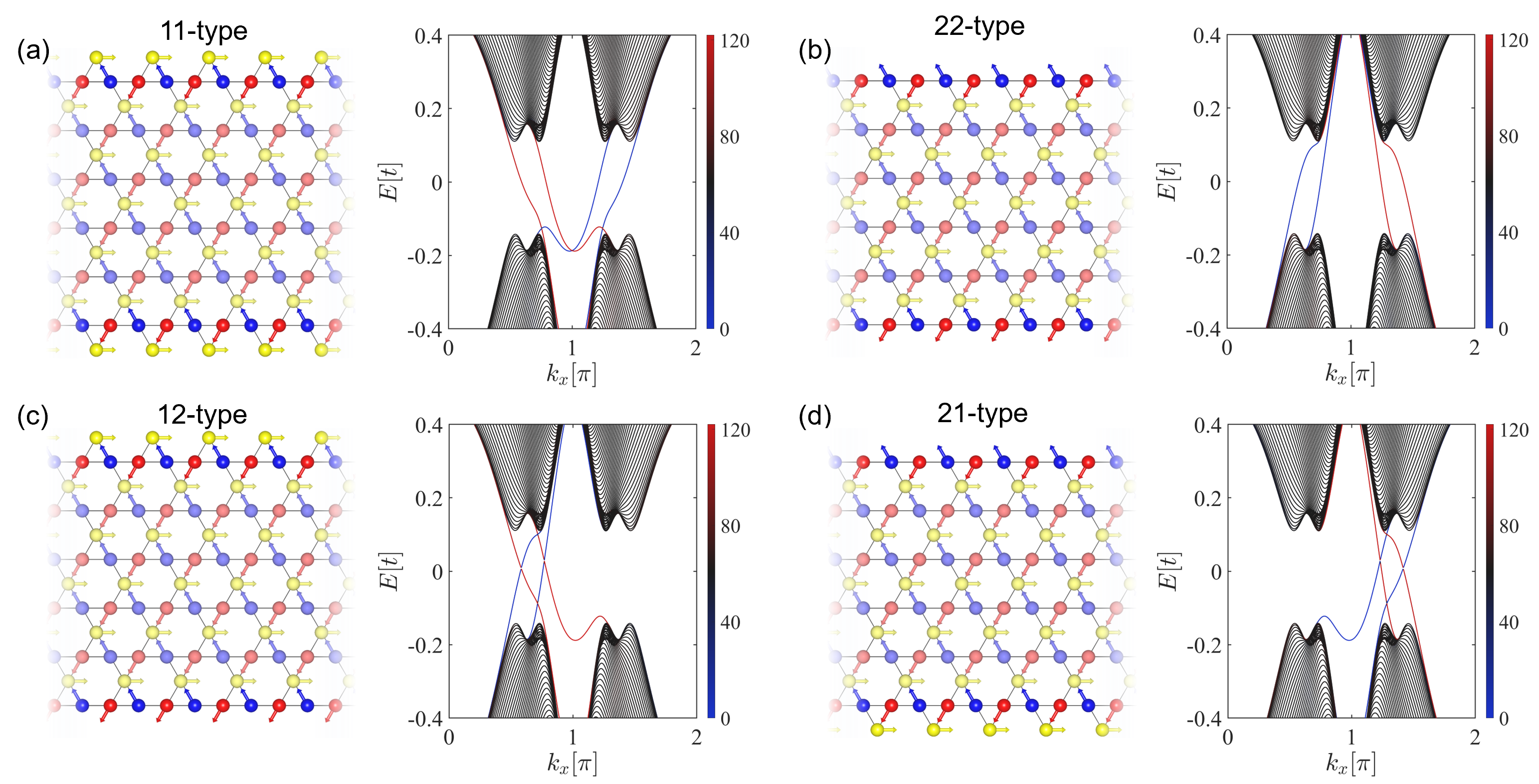}
\caption{(a) Left: Schematic of a chiral kagome antiferromagnet ribbon with 11-type boundary termination  in the $y$ direction (i.e., where both the upper and lower edges terminate on one sublattice). Right: Energy spectrum for $J=0.6 t$ and $M_z=0.2 t$. The colorbar represents the expectation value of the position operator $\langle \hat{y} \rangle$ (in units of $\sqrt{3}a/2$). 
(b-d) Same as (a) but for the 22-, 12- and 21-type boundary terminations, respectively. 
}
\label{fig:Band_OpenY}
\end{figure*}

As a concrete realization, we consider chiral kagome antiferromagnets. Chiral kagome antiferromagnets exhibit spin-split electronic structures despite having vanishing net magnetization~\cite{Moessner01CJP,Matan06PRL,Chen2014PRL,Kubler14EPL}. Representative materials include Mn$_3$Sn, Mn$_3$Ge and etc, which have attracted major interest because of their rich topological and transport properties  exhibit~\cite{Nakatsuji15Nature,Nayak16SciAdv,Kuroda17NatMater,Zelezny17PRL,Ikhlas17NatPhys,Higo18NatPhotonics,tsai2020electrical,higo2022perpendicular,zhou2026field}.
We show that out-of-plane spin canting or spin–orbit coupling opens a topological bulk gap with a high Chern number, giving rise to multiple chiral edge states (Fig.~\ref{fig:Lattice_Band}). Although the chirality of these edge modes is fixed by topology, their valley location and dispersion are remarkably sensitive to sublattice-dependent boundary termination, producing asymmetric edge spectra while leaving the bulk electronic structure essentially unchanged. When proximity-coupled to conventional superconductors, the resulting transverse boundary asymmetry is converted into longitudinal nonreciprocal superconducting transport, leading to asymmetric Andreev bound-state spectra, nonreciprocal Josephson currents, and anomalous Fraunhofer interference patterns.
This mechanism is qualitatively distinct from previous proposals based on time-reversal-symmetric helical edge states~\cite{Chen2018PRB,Souto2022PRL,YPLi24NC,Huang24APL,GuoGL25arXiv}, where the diode effect requires a finite magnetic flux and can be understood as interference between two localized supercurrents. Instead, our results show that transverse boundary asymmetry alone can govern longitudinal superconducting nonreciprocity, establishing a boundary-controlled route to the Josephson diode effect without relying on bulk electronic asymmetry.

\textit{\color{blue}Chern gap with $\mathcal{C}=2$}.\textemdash
To elucidate the essential physics, we consider a prototypical kagome-lattice model for chiral antiferromagnets with a coplanar 120$^\circ$ spin texture [Fig.~\ref{fig:Lattice_Band}(a)].
The lattice consists of corner-sharing triangles with three sublattices, $A$, $B$, and $C$. The Hamiltonian reads~\cite{Chen2014PRL}
\begin{equation}
H_{\text{cAFM}} = -t\sum_{\langle \bm{r}_i\alpha,\bm{r}_j\beta\rangle} c_{\bm{r}_i\alpha}^\dagger c_{\bm{r}_j\beta}
+ J\sum_{\bm{r}_i\nu} c_{\bm{r}_i\nu}^\dagger \bm{s}\cdot \bm{S}_{\bm{r}_i\nu} c_{\bm{r}_i\nu} ,
\end{equation}
where \(t\) denotes the nearest-neighbor hopping amplitude,  \(J\) the exchange coupling, $c_{\bm{r}_i\alpha}=(c_{\bm{r}_i\alpha\uparrow},c_{\bm{r}_i\alpha\downarrow})^T$ the electron spinor on sublattice $\alpha$ at site $\bm{r}_i$, \(\bm{s}=(s_x,\,s_y,\,s_z)\) is the electron spin operator, and $\bm{S}_{\bm{r}_i\nu}=(\cos\theta_\nu, \sin\theta_\nu, 0)$ represents the local exchange field on sublattice $\nu \in \{A,\,B,\,C\}$. Throughout this work, we adopt $(\theta_A,\theta_B,\theta_C) =(2\pi/3,-2\pi/3,0)$ in our calculation. Other choices related by a global spin rotation or chirality reversal do not affect the main results up to sign reversals of the topological responses.

The coplanar magnetic order breaks time-reversal symmetry $\mathcal{T}$, but preserves the combined symmetry $\mathcal{TM}_z$, where $\mathcal{M}_z$ is mirror reflection about the kagome plane. The bulk model also preserves inversion symmetry $\mathcal{I}$. The $\mathcal{TM}_z$ symmetry imposes $\Omega_z({\bf k})=-\Omega_z(-{\bf k})$, while inversion gives $\Omega_z({\bf k})=\Omega_z(-{\bf k})$. Thus, the Berry curvature vanishes $\Omega_z({\bf k})= 0$, thereby forbidding a finite Chern number. Accordingly, in the band strcture, symmetry-protected degeneracies appear at the $\Gamma$ and $K(K')$ points~\cite{Chen2014PRL,ZhangSB25PRB,zhang2026Newton} [gray dotted curves in Fig.~\ref{fig:Lattice_Band}(b)]. 

To open a topological bulk gap, the $\mathcal{TM}_z$ symmetry must be broken. This can be achieved, for example, by an out-of-plane spin canting described by an exchange term 
\begin{align}
   H_{\text{cant}} =  M_z\sum_{\bm{r}_i} c^\dagger_{\bm{r}_i}\, s_z\, c_{\bm{r}_i} 
\end{align}
or by spin-orbit coupling as discussed in the SM~\cite{sm2026_edge_cAFM_diode}. Both approaches remove the symmetry constraint on the Berry curvature and can drive the system into a Chern-insulating phase, although their microscopic origins differ. For the sake of concreteness, we focus in the following on the spin canting and present the results of the spin-orbit coupling case in the SM. In the spin-canting case, $M_z$ acts as a \(\mathcal{TM}_z\)-breaking mass term for the low-energy states near $K$ and $K'$, and the magnetic texture becomes essentially non-coplanar.  
The gaped phase can be characterized by the Chern number of the occupied bands~\cite{Thouless82PRL},
\begin{equation}
\mathcal C
= \sum_{n\in \mathrm{occ}} \mathcal{C}_n, \quad  \mathcal{C}_{n}=
\frac{1}{2\pi}
\int_{\mathrm{BZ}} d^2 k\, \Omega_{n,z}(\bm{k}).
\end{equation}
Finite $M_z$ separates all the bands and endows them with nonzero Chern numbers, as shown in Fig.~\ref{fig:Lattice_Band}(b). At 1/3 filling, a bulk gap with Chern number \(\mathcal C=\pm 2\) is obtained, with the sign determined by the direction
of $M_z$. The Berry curvature is concentrated near the gaped valleys. According to the bulk-boundary correspondence~\cite{Hatsugai93PRL,Shen2012book}, this gap supports two co-propagating chiral edge states under open boundary conditions, with their low-energy spectral weight located near the valleys.

\textit{\color{blue}Asymmetric and valley-dependent chiral edge states}.-- 
While the Chern number fixes the chirality of edge states, it does not uniquely determine their valley character, spatial distribution and detailed dispersion. These properties are instead controlled by the boundary termination. 
To illustrate this point, we consider a ribbon geometry that is open in the $y$ direction and translationally invariant along $x$. Since the kagome lattice contains three sublattices, different cuts expose different sublattice terminations at two edges. We label the boundary type by the number of sublattices exposed at the two edges. This gives two distinct classes: symmetric terminations, where both edges terminate with one sublattice $C$ for the 11-type case or two sublattices $A/B$ for the 22-type case [Fig.~\ref{fig:Band_OpenY}], and asymmetric terminations, where one edge terminates with two sublattices $A/B$ and the opposite edge terminates with one sublattice $C$ for the 12- and 21-type cases. 

Figure~\ref{fig:Band_OpenY} shows the ribbon spectra for different boundary terminations. For the symmetric terminations (11/22-type), the upper and lower edges have the same sublattice termination and can be related by the spatial mirror symmetry about the ribbon centerline. Consequently, the edge modes localized at opposite boundaries form symmetry-related pairs, yielding edge-resolved spectra that satisfy $E_{\text{upper}}(k_x)=E_{\text{lower}}(-k_x)$, as illustrated in Figs.~\ref{fig:Band_OpenY}(a,b). This leads to balanced low-energy contributions from the two edges.

The situation changes qualitatively for the asymmetric terminations (12/21-type). Because the two edges expose different sublattice configurations, the chiral states at opposite edges are no longer symmetry-related. Consequently, the relation $E_{\text{upper}}(k_x)=E_{\text{lower}}(-k_x)$ breaks down, and the edge modes at the two boundaries develop distinct dispersions, as shown in Figs.~\ref{fig:Band_OpenY}(c,d). 
Moreover, the low-energy edge modes at both boundaries are selected near the same valley while propagating in opposite directions, producing a pronounced valley-selective edge response. Thus, boundary asymmetry generates inequivalent edge spectra at opposite boundaries, even though the bulk spectrum remains symmetric about $k_x=0$.

The microscopic origin of this behavior lies in the sublattice-valley structure of the kagome bands. Near $K$ and $K'$, the low-energy wavefunctions carry distinct sublattice characters. This allows edge termination to act as a valley-selective filter by favoring the valley associated with the exposed sublattices. For chiral modes localized at a given edge, $A/B$- and $C$-terminated edges favor opposite valleys. This explains why changing the boundary termination modifies both the dispersion and valley location of the low-energy edge modes. The valley character is further controlled by the spin canting $M_z$: Reversing $M_z$ flips the sign of Chern number and the edge-state chirality, while transferring the spectral weight between $K$ and $K'$. Thus, the propagation direction, valley character, and boundary localization of the chiral states are jointly controlled by the band topology, spin canting, and sublattice termination.

\begin{figure}[t]
\centering
\includegraphics[width=1\linewidth]{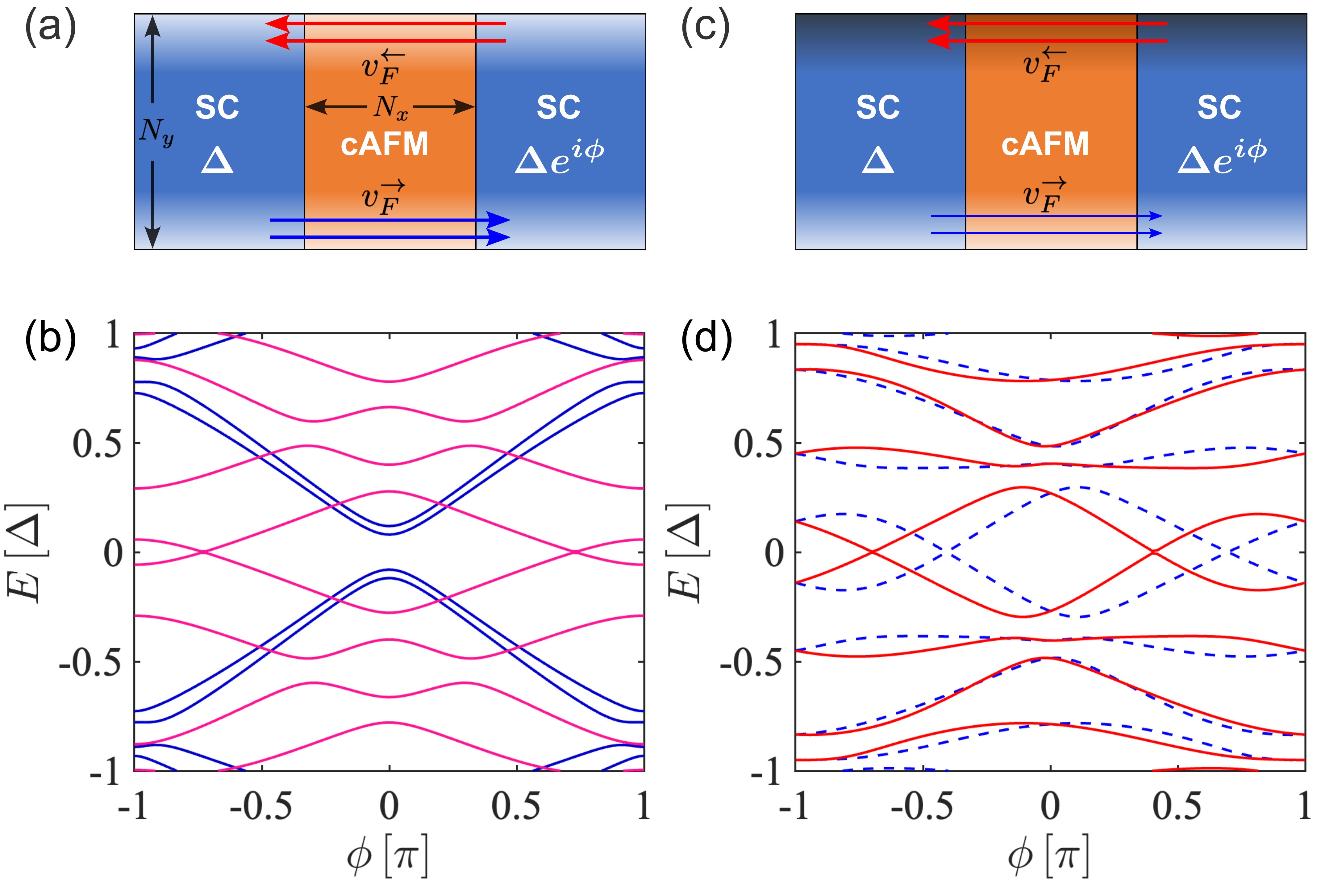}
\caption{
Andreev spectrum of an SNS junction with a chiral antiferromagnetic spacer under an out-of-plane spin canting $M_z$. 
(a) Schematic of a junction oriented along the $x$ direction with symmetric transverse boundaries, corresponding to the 11- or 22-type termination. The cAFM spacer has length $N_x$ (in units of lattice constant $a$) in the transport ($x$) direction and width $N_y$ (in units of $\sqrt{3}a$) in the transverse ($y$) direction. 
(b) Low-energy Andreev spectrum as a function of superconducting phase difference $\phi$. The pink and blue curves correspond to the 11- and 22-type terminations. 
(c,d) Same as (a,b) but for the asymmetric (12-type) boundary configuration. 
Solid red and dashed blue curves represent $E(\phi)$ and $E(-\phi)$, respectively, highlighting the asymmetry of the spectrum under $\phi\to -\phi$. Other parameters are $J=0.4t$, $M_z=0.1t$, $\mu=-0.1t$, $N_x=N_y=10$, $\Delta=0.03t$ and $T=0.02\Delta/k_B$. 
}
\label{fig:ABS_Bz}
\end{figure}

\textit{\color{blue}Asymmetric Andreev spectrum}.\textemdash The asymmetric chiral edge states have direct signatures in the Andreev bound-state spectra when coupled to superconducting leads. For illustration, we consider an SNS junction along the $x$ direction, where the canted chiral antiferromagnet forms the normal region and two ordinary $s$-wave superconducting leads have the same pairing potential $\Delta$ but a phase difference $\phi$, as schematically shown in Figs.~\ref{fig:ABS_Bz}(a,c). We set the Fermi level inside the bulk gap of the antiferromagnet, such that the low-energy transport is dominated by the chiral edge modes. The resulting Andreev spectra are shown in Figs.~\ref{fig:ABS_Bz}(b,d). Details of the calculation are provided in the SM~\cite{sm2026_edge_cAFM_diode}. 

For symmetric transverse boundaries (11/22-type), the chiral modes localized at the two opposite edges are equivalent. In this case, reversing the phase difference $\phi$ effectively maps an Andreev trajectory on one edge to a symmetry-related trajectory on the opposite edge. Since the two trajectories experience equivalent boundary environments and Fermi velocities, the total phases accumulated during the Andreev cycles remain balanced. Consequently, the Andreev spectrum is symmetric under phase reversal, $E(\phi)=E(-\phi)$,   as shown in Fig.~\ref{fig:ABS_Bz}(b).

In contrast, for the asymmetric 12-type termination shown in Fig.~\ref{fig:ABS_Bz}(c), the two edges are no longer symmetry-related, and the corresponding chiral edge modes have different valley characters and Fermi velocities, as discussed above. These differences modify the propagation phases accumulated by the electron and hole components during an Andreev cycle. The two edge trajectories therefore acquire unequal effective phase shifts when coupled to the superconducting leads. Under $\phi\rightarrow -\phi$,
an Andreev trajectory is no longer mapped onto an equivalent trajectory at the opposite edge, removing the symmetry constraint that enforces a phase-symmetric spectrum. The calculated Andreev spectrum therefore becomes phase asymmetric, $E(\phi)\neq E(-\phi)$, as shown in Fig.~\ref{fig:ABS_Bz}(d). This asymmetry directly reflects the boundary-induced inequivalence of the chiral edge states.

\begin{figure}[t]
\centering
\includegraphics[width=1\linewidth]{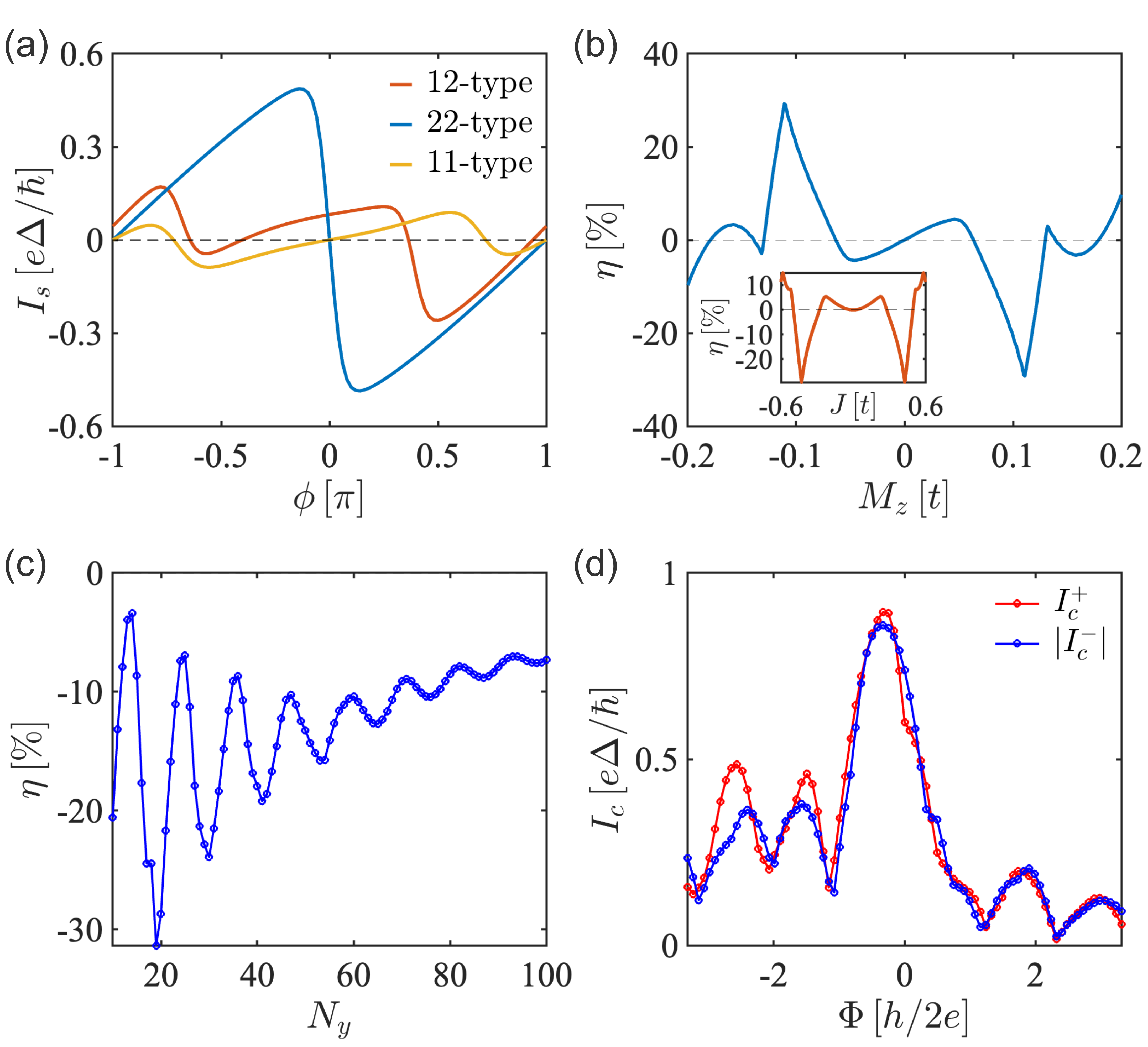}
\caption{(a) CPR of the SNS junction for different transverse boundary terminations, with $J=0.4t$, $M_z=0.1t$ and $N_y=10$. The asymmetric 12-type termination gives rise to an asymmetric CPR, yielding unequal forward and backward critical currents, $I_c^+\neq |I_c^-|$, and a diode efficiency $\eta=-20.58\%$. (b) Diode efficiency $\eta$ as a function of $M_z$ for $J=0.4t$ and $N_y=10$. Inset: $\eta$ as a function of $J$ for $M_z=0.1t$ and $N_y=10$. (c) $\eta$ as a function of $N_y$ for $J=0.4t$ and $M_z=0.1t$. (d) Direction-dependent Fraunhofer pattern, showing $I_c^+$ and $|I_c^-|$ as functions of magnetic flux $\Phi$ (in units of $h/2e$) for $J=0.4t$, $M_z=0.1t$ and $N_y=60$. Other parameters are $\mu=-0.1t$, $N_x=10$, $\Delta=0.03t$ and $T=0.02\Delta/k_B$. (b-d) are calculated for $12$-type transverse boundary termination.
}
\label{fig:Diode}
\end{figure}

\textit{\color{blue}Nonreciprocal supercurrents}.\textemdash The asymmetric chiral edge states also convert transverse boundary asymmetry into longitudinal Josephson nonreciprocity, despite an inversion-symmetric bulk electronic structure. This contrasts with prevailing superconducting diode mechanisms based on bulk-state asymmetry~\cite{HeJ2022NJP,Yuan2022PNAS,ZhangY2022PRX}. We calculate the supercurrent employing the Green function technique described in the SM~\cite{sm2026_edge_cAFM_diode}. Figure~\ref{fig:Diode} shows representative current-phase relations (CPRs) for different boundary terminations. 

For the symmetric 11/22-type terminations, the edge channels at opposite boundaries remain balanced. Consequently, the CPR is reciprocal, $I(\phi)=-I(-\phi)$, leading to equal forward and backward critical currents, $I_c^+=|I_c^-|$. 
In contrast, for the asymmetric 12-type termination, the phase-asymmetric Andreev spectrum gives rise to an anomalous CPR with a shifted free-energy minimum, characteristic of a $\phi_0$ junction. Importantly, the inequivalent edge contributions not only shift the CPR but also distort its shape, leading to unequal critical currents, $I_c^+ \neq |I_c^-|$ [red curve in Fig.~\ref{fig:Diode}(a)]. 
The contrast between the 12/21-type and 11/22-type terminations demonstrates that transverse boundary asymmetry is essential for generating the longitudinal diode response.

The chiral-edge-state mechanism imposes a field-reversal relation on the Josephson current. We define the diode efficiency as
$\eta =({I_c^+ - |I_c^-|})/({I_c^+ + |I_c^-|})$.  The supercurrent satisfies $I_s(M_z,\phi)=-I_s(-M_z,-\phi)$~\cite{sm2026_edge_cAFM_diode}, which implies $\eta(M_z)=-\eta(-M_z)$ [Fig.~\ref{fig:Diode}(b)]. Thus, reversing $M_z$ reverses the chiral-edge asymmetry and switches the diode polarity. Moreover, $\eta$ evolves continuously with $M_z$ and remains finite for small canting. This indicates that the diode effect can persist even when bulk states contribute to transport. The inset further shows that $\eta$ is even in $J$ but vanishes at $J=0$, confirming that both chiral magnetic order and boundary asymmetry are essential for the diode effect. 
The width $N_y$ dependence in Fig.~\ref{fig:Diode}(c) further reflects the edge-state origin. 
For small $N_y$, $\eta$ exhibits pronounced oscillations due to hybridization between opposite edge channels.
As $N_y$ increases, this finite-size coupling becomes weaker, the oscillations gradually weaken and $\eta$ approaches a saturated value. 

\textit{\color{blue}Anomalous Fraunhofer patterns}.\textemdash
This mechanism also gives rise to anomalous Fraunhofer patterns under a perpendicular magnetic flux $\Phi$ threading the junction. In a conventional reciprocal junction with a uniform current distribution, magnetic flux produces the standard sinc-like Fraunhofer interference pattern, for which $I_c^+(\Phi)=|I_c^-(\Phi)|$ and the principal maximum occurs at \(\Phi=0\). In contrast, in the present setup, the forward and backward supercurrents are carried predominantly by inequivalent edge channels at opposite edges. The applied flux therefore imprints different phase shifts on spatially separated edge trajectories, whose amplitudes, dispersions, and effective phase offsets are boundary dependent. As a result, the forward and backward critical currents exhibit direction-dependent interference patterns: $I_c^+(\Phi)$ and $|I_c^-(\Phi)|$ differ not only in magnitude, but also in their lobe structures and peak positions, as shown in Fig.~\ref{fig:Diode}(d). The principal maxima are shifted away from $\Phi=0$, accompanied by strongly asymmetric side-lobe structures. These features are consistent with the anomalous Josephson character shown in Fig.~\ref{fig:Diode}(a), but reflecting inequivalent spatial distributions of forward and backward edge supercurrents rather than a simple uniform $\phi_0$ offset. 
The direction-dependent Fraunhofer patterns therefore provide another experimentally accessible and distinct signature of boundary-controlled chiral edge states.

\textit{\color{blue}Discussion}.\textemdash
We proposed a boundary-controlled mechanism for nonreciprocal superconducting transport based on chiral edge states. In this mechanism, transverse boundary asymmetry makes the chiral edge channels at opposite edges inequivalent, thereby generating a longitudinal nonreciprocal response even when the bulk preserves inversion symmetry. Using chiral kagome antiferromagnets as an example, we showed that out-of-plane spin canting realizes a high-Chern-number phase with multiple chiral edge modes whose dispersions are strongly controlled by boundary termination. This boundary-controlled edge inequivalence manifests directly in asymmetric Andreev spectra, nonreciprocal critical supercurrents, and anomalous Fraunhofer interference patterns. 

This mechanism is generic and does not rely on fine tuning. As shown in the SM~\cite{sm2026_edge_cAFM_diode}, similar boundary-controlled asymmetric chiral edge states and nonreciprocal transport appear for junctions oriented along the $y$ direction, for the topological gap near 5/6 filling, and for spin-orbit-coupling-induced topological phases. The diode effect also persists when bulk states contribute to transport, indicating that a purely edge-only regime is not required. While the above analysis focused on regular asymmetric 12/21 boundary terminations, similar behavior is expected for more general asymmetric or irregular boundaries that render the opposite edge channels inequivalent, as naturally expected in realistic devices with imperfect or engineered edges. 
More broadly, this principle can be extended to other Chern-insulating magnetic systems, including bilayer or multilayer structures with engineered boundary-dependent chiral channels. 

Our predictions may be tested in candidate materials, including the experimentally established chiral kagome antiferromagnets Mn$_3$Sn, Mn$_3$Ge, and related compounds~\cite{tsai2020electrical,takeuchi2021chiral,higo2022perpendicular,xie2022magnetization,pal2022setting,zheng2024effective,zheng2025all,xu2025spin,zhou2026field}. Notably, thin-film Josephson junctions based on these materials are becoming experimentally accessible~\cite{jeon2021long,jeon2023chiral}, while controlled boundary terminations may be engineered by nanofabrication, edge-selective growth, or patterned device geometries~\cite{jia2009controlled,stanford2018emerging,duan2022epitaxial,mazzola2023observation,wei2026scalable}. This mechanism therefore highlights the engineering of opposite boundaries as a feasible route toward nonreciprocal superconducting devices.

\begin{acknowledgments}
We thank Lunhui Hu, Chuang Li, and Zhenyu Zhang for valuable discussions. 
This work was supported by the National Natural Science Foundation of China (Grant No. 12488101), HFNL Self-Deployed Project (Grant No. ZB2602000300), and Quantum Science and Technology-National Science and Technology Major Project (Grant No. 2021ZD0302801). 
\end{acknowledgments}

\bibliography{Ref_edge_diode}

\end{document}